\documentclass[cits]{PoS}
\usepackage{xspace}
\usepackage{amssymb}
\usepackage{fontenc}
\usepackage{times}
\usepackage{mathptmx}
\usepackage{graphicx}
\usepackage{natbib}

\usepackage{epsfig}

\def \etal{et al.\xspace}

\title{A dense gas survey of the gamma-ray sources HESS\,J1731$-$347 and HESS\,J1729$-$345 }

\ShortTitle{A dense gas survey of HESS\,J1731$-$347 and HESS\,J1729$-$345 }

\author{\speaker{Nigel Maxted}$^a$, Gavin Rowell$^b$, Phoebe de Wilt$^b$, Michael Burton$^c$, Matthieu Renaud$^a$, Yasuo Fukui$^d$, Jarryd Hawkes$^b$, Rebecca Blackwell$^b$, Fabien Voisin$^b$, Vicki Lowe$^{ce}$ and Felix Aharonian$^f$\\
        \llap{$^a$}Laboratoire Univers et Particules de Montpellier, Universite de Montpellier 2, France\\
        \llap{$^b$}School of Chemistry and Physics, University of Adelaide, Adelaide, Australia\\
        \llap{$^c$}School of Physics, University of New South Wales, Sydney, Australia\\
        \llap{$^d$}Department of Astrophysics, Nagoya University, Japan\\
        \llap{$^e$}Australia Telescope National Facility, CSIRO Astronomy and Space Science, Australia\\
        \llap{$^f$}Max-Planck-Institut fur Kernphysik, Heidelberg, Germany\\
        E-mail: \email{nigel.maxted@univ-montp2.fr}}
           
\abstract{The results of Mopra molecular spectral line observations towards the supernova remnant HESS\,J1731$-$347 (G353.6$-$0.7) and the unidentified gamma-ray source HESS\,J1729$-$345 are presented. Dense molecular gas in three different velocity-bands (corresponding to three Galactic arms) are investigated using the CS(1-0) line. The CS-traced component provides information about the dense target material in a hadronic scenario for gamma-ray production (cosmic-rays interacting with gas) and an understanding of the dynamics. Furthermore, the effects of cosmic ray diffusion into dense gas may alter the gamma-ray spectrum to cause a flattening of spectra towards such regions. Dense molecular gas mass at a level of $\sim$10$^5$\,M$_{\odot}$ was revealed in this survey, with mass of the order of $\sim$10$^3$\,M$_{\odot}$ towards HESS\,J1729$-$345 in each coincident Galactic arm, but no significant detection of dense molecular gas towards HESS\,J1731$-$347 at the currently-preferred distance of $\sim$5.2-6.2\,kpc was discovered.
}

\FullConference{Cosmic Rays and the InterStellar Medium\\
                 24-27 June 2014\\
                 Montpellier, France}

\begin{document}

\section{Introduction}
\begin{sloppypar}
HESS\,J1731$-$347 is a TeV gamma-ray source \citep{Aharonian:2008,Abramowski:2011} associated with the shell-type supernova remnant (SNR), G353.6$-$0.7 \citep{Tian:2008}. The nature of the nearby gamma-ray source HESS\,J1729$-$345 (north of HESS\,J1731$-$347, see Figure\ref{Fig:Master}) has remained unclear since being resolved as a separate TeV gamma-ray source \citep{Abramowski:2011} and a counterpart is yet to be identified at other wavelengths.
\end{sloppypar}
HESS\,J1731$-$347 is known to be accelerating leptons beyond TeV energies due to the detection of non-thermal X-rays \citep{Tian:2008,Tian:2010,Bamba:2012}. Thus, like IC\,443 and W44 \citep{Ackermann:2013}, HESS\,J1731$-$347 may plausibly also accelerate hadrons beyond TeV energies. Indeed, gamma-ray emission like that seen with HESS can be produced from cosmic-ray (CR) hadron interactions with molecular gas (hadronic production), so knowledge of the gas distribution may help to distinguish between this and a competing mechanism for gamma-ray production, inverse Compton scattering of photons by electrons (leptonic production).

\citet{Fukuda:2014} argue that the HESS\,J1731$-$347 SNR is associated with a void in atomic 3\,kpc-expanding Arm gas at a distance of 5.2-6.2\,kpc (line of sight velocity $\sim - 85$\,km\,s$^{-1}$). Their analysis of the surrounding HI and CO emission could plausibly identify a large proportion of the gas-mass required for a hadronic gamma-ray production mechanism to be consistent with observations, but the authors argue that such a scenario would still include a $\sim$20\% leptonic contribution to the total gamma-ray flux (much of this in the southern region of the remnant).

Certainly, upper limits placed on GeV emission using Fermi-LAT data \citep{Yang:2014} seem to disfavour a hadronic model for gamma-ray production. 
However, a hadronic model may still be plausible given the potential effects of cosmic ray hadron diffusion into dense gas clumps, as suggested by \citet{Fukuda:2014}. In this model, the highest energy cosmic rays have access to a larger amount of target material (e.g. \citealt{Gabici:2009}).

We used the Mopra radio telescope to survey the HESS\,J1731$-$347 (and HESS\,J1729$-$345) region at 7\,mm wavelengths, targeting the CS(1-0) line to trace dense molecular gas which may play a role in a hadronic gamma-ray production scenario. We simultaneously recorded emissions of several other warm gas tracers.

\section{Observations \& Analysis}
The field displayed in Figure\,\ref{Fig:Master} was surveyed in March 2011, April 2012, November 2013 and May 2014 using the 22\,m Mopra radio telescope, located $\sim$450\,km northwest of Sydney, Australia (31$^{\circ}$16$^{\prime}$04$^{\prime\prime}$S, 149$^{\circ}$05$^{\prime}$59$^{\prime\prime}$E). The 7\,mm spectrometer configuration and data reduction techniques were the same as those presented in \citet{Maxted:2013b}.

Mopra mapping data have a spacing between scan rows of 26$^{\prime\prime}$. The velocity resolution of the 7\,mm zoom-mode data is $\sim$0.2\,kms$^{-1}$. The beam FWHM and the pointing accuracy of Mopra at 7\,mm are 59$\pm$2$^{\prime\prime}$ and $\sim$6$^{\prime\prime}$, respectively \citep{Urquhart:2010}. The Mopra spectrometer, MOPS, is capable of recording sixteen tunable, 4096-channel (137.5\,MHz) bands simultaneously when in `zoom' mode, as used here.

The exposure varies over the mapped field, with 1$\sigma$ noise levels generally within the range 0.04-0.08\,K, but exceeding 0.1\,K for $\sim$1\% of map pixels due to bad weather. To adjust for this, a conservative threshold of 4.5$\sigma$, calculated separately for each pixel, was placed on maps to ensure false detections were not included in images.

Column densities were generated from CS(1-0) maps by applying the procedure presented in \citet{Maxted:2012a}. Eq.\,9 of \citet{Goldsmith:1999} was used to convert CS(1-0) integrated emission into CS(J=1) column density. CS(1-0) emission was assumed to be optically thin unless a detection of C$^{34}$S(1-0) emission was present, in which case optical depth was estimated assuming a CS-C$^{34}$S ratio of 22.5. To convert CS(J=1) column density into total CS column density and subsequently H$_2$ column density, a temperature of 20\,K and a CS abundance relative to H$_2$ of 10$^{-9}$ \citep{Frerking:1980} were assumed, respectively. Systematic effects due to these two assumptions are expected to be the largest contributor to the uncertainty in results, with molecular abundance possibly varying by an order of magnitude (\citealt{Maxted:2012a} and references therein). Column density images of regions where optical depth was calculated were smoothed by a Gaussian with a width of 2$^{\prime}$ (two 7\,mm Mopra beam-widths) to smooth the boundary between optically thick and thin analyses.

\section{Results \& Discussion}
\begin{figure}
\includegraphics[width=.99\textwidth]{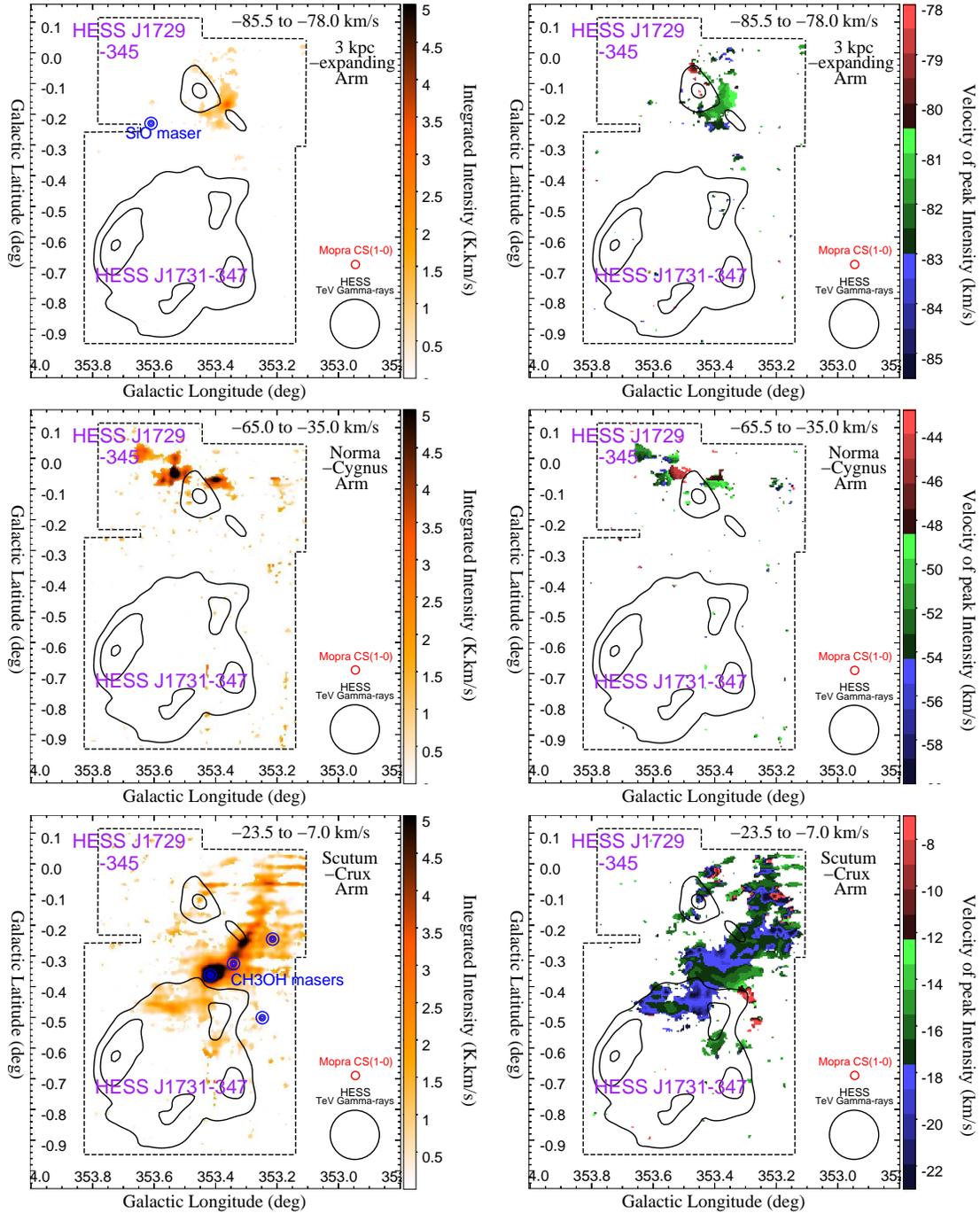}
\caption{Mopra CS(1-0) integrated intensity (left-side) and velocity centroid (right-side) images of 3 velocity-intervals (range indicated) of CS(1-0) emission, with overlaid 4, 6 and 8$\sigma$ HESS TeV gamma-ray emission contours \citep{Abramowski:2011}. The Mopra CS(1-0) and HESS TeV gamma-ray beam sizes are indicated in each picture. Exposure varies across the field (represented by the dotted line), so these images have been thresholded at a non-constant level corresponding to a signal-noise ratio of 4.5.}
\label{Fig:Master}
\end{figure}

As displayed in Figure\,\ref{Fig:Master}, CS(1-0) clumps were discovered at three velocity ranges towards the HESS\,J1731$-$347/HESS\,J1729$-$345 field corresponding to three Galactic arms: the 3\,kpc-expanding arm, the Norma-Cygnus arm and the Scutum-Crux arm. The critical density for emission of the CS(1-0) line is $\sim$10$^4$\,cm$^{-3}$, so Figure\,\ref{Fig:Master} highlights regions containing gas of a similar or higher density.


\subsection{3\,kpc-expanding Arm}
Dense gas traced by CS(1-0) emission at $\sim$6\,kpc (line of sight velocity $\sim -85$\,km\,s$^{-1}$, top of Figure\,\ref{Fig:Master}) probably corresponds to the 3\,kpc-expanding arm \citep{Vallee:2008,Fukuda:2014}. HESS\,J1731$-$347 is believed to be associated with gas at this distance/velocity, however no link has been established between the northern source HESS\,J1729$-$345, and either HESS\,J1731$-$347 or gas at this velocity. Dense gas lies outside the border of HESS\,J1729$-$345 and although $\sim$4$\times$10$^4$\,M$_{\odot}$ was traced at this velocity, only $\sim$10$^3$\,M$_{\odot}$ is coincident with the gamma-ray emission (at a 4$\sigma$ level).

\begin{sloppypar}
Although $\sim$6.4$\times$10$^4$\,M$_{\odot}$ of atomic and molecular gas was previously traced towards HESS\,J1731$-$347 using the HI and CO(1-0) lines \citep{Fukuda:2014}, CS(1-0) emission was not detected towards this source, suggesting an absence of dense ($\sim$10$^4$\,cm$^{-3}$) gas detectable at a level above antenna intensity 0.04-0.08\,K. This may disfavour (although not strictly rule-out) the notion that clumps of dense gas are having an effect on the high energy gamma-ray spectrum in a hadronic scenario. Specifically, diffusion into dense gas has been suggested to allow the highest energy cosmic ray hadrons to access more target material than their lower energy counterparts in the case of RX\,J1713.7$-$3946 \citep{Zirakashvili:2010,Inoue:2012,Maxted:2012a} and such scenarios should be investigated further for HESS\,J1731$-$347. We note that so-called `dark' molecular gas \citep{Wolfire:2010} may exist towards the region (untraced by HI, CO and CS) and this may also affect the gamma-ray spectrum.
\end{sloppypar}

\begin{sloppypar}
We also note the Mopra detection of SiO(1-0,v=2) emission at line of sight velocity $\sim - 85$\,km\,s$^{-1}$ towards the location of the IR star OH\,353.61$-$0.23 \citep{Sevenster:1997}, which is not expected to be related to either of the HESS TeV gamma-ray sources.
\end{sloppypar}

\subsection{Norma-Cygnus Arm}
\begin{sloppypar}
Dense gas traced by CS(1-0) emission at $\sim$5\,kpc (line of sight velocity $\sim - 55$\,km\,s$^{-1}$, centre images of Figure\,\ref{Fig:Master}) probably corresponds to the Norma-Cygnus arm \citep{Vallee:2008}. Although CS(1-0) emission traces $\sim$10$^5$\,M$_{\odot}$ of dense gas, no significant clumps are coincident with HESS\,J1731$-$347, while $\sim$5$\times$10$^3$\,M$_{\odot}$ are coincident with HESS\,J1729$-$345 gamma-ray emission (at a 4$\sigma$ level).
\end{sloppypar}

\subsection{Scutum-Crux Arm}
Dense gas at $\sim$3.2kpc (v$_{LSR} \sim -23.5\rightarrow-$7.0\,km\,s$^{-1}$, bottom of Figure\,\ref{Fig:Master}) may be a component of the Scutum-Crux arm \citep{Vallee:2008,Abramowski:2011}. This CS(1-0) clump harbours the HII region, G353.42$-$0.37, which was exploited to place a lower limit of $\sim$3.2\,kpc on the HESS\,J1731$-$347 SNR in previous absorption studies \citep{Tian:2008,Abramowski:2011}. Towards G353.42$-$0.37 Mopra recorded detections of C$^{34}$S(1-0), CH$_3$OH(7-6), HC$_3$N(5-4) and SiO(1-0) emission at velocities consistent with the CS(1-0) emission, suggesting a warm, dense and shocked environment such as that commonly associated with star-formation or gas irradiated/warmed by stellar activity. Three other CH$_3$OH(7-6) detections were also detected (see bottom of Figure\,\ref{Fig:Master}). Two at 353.34$^{\circ}-$0.33$^{\circ}$ and 353.21$^{\circ}-$0.25$^{\circ}$ have no clear counterparts, but a detection at 353.25$^{\circ}-$0.50 possibly corresponds to a millimetre continuum source (BGPS\,G353.216-00.246, \citealt{Rosolowsky:2010}). We note that the velocities of the four CH$_3$OH(7-6) emission lines match the velocities of the coincident CS(1-0) clumps (see Figure\,\ref{Fig:Master}, details to be presented in a later publication). The dense gas has a varied velocity structure between $-$23.5 and $-$7\,km\,s$^{-1}$, suggesting that it may be either composed of clumps with a range of local velocities or clumps spatially-distinct from one another (with the distance reflected in rotation velocity). No clear correspondence between velocity features and gamma-ray sources are observed.

\begin{sloppypar}
Analyses of the CS(1-0) emission line reveal $\sim$5$\times$10$^5$\,M$_{\odot}$ of dense molecular gas, $\sim$5$\times$10$^4$\,M$_{\odot}$ of which overlaps the northern half of HESS\,J1731$-$347, while $\sim$6$\times$10$^3$\,M$_{\odot}$ overlaps HESS\,J1729$-$345. 
\end{sloppypar}




\section{Summary and Future Work}
We detect dense molecular gas towards the HESS\,J1731$-$347/HESS\,J1729$-$345 region using the CS(1-0) emission line, but we do not detect CS(1-0) emission coincident with HESS\,J1731$-$347 at the currently-favoured velocity ($\sim - 85$\,km\,s$^{-1}$) in this survey.
We discover dense molecular gas towards and around HESS\,J1729$-$345 at three separate velocities. Despite the existence of coincident masses of $\sim$10$^3$\,M$_\odot$, we are unable to identify an association between the gas and the HESS\,J1729$-$345 gamma-ray emission.



\end{document}